\documentclass[twoside,openright,a4paper]{article}
\usepackage{graphicx}
\usepackage{graphics}
\usepackage{epsfig}
\usepackage{amsmath}
\usepackage{amsfonts}
\usepackage{amssymb}
\usepackage[version=3]{mhchem}
\usepackage{sectsty}
\sectionfont{\large}
.360pk
\font\rt=cmss9.360pk
\font\sd=cmcsc9.360pk

\begin{document}

\begin{center}\Large 3D Simulation of Dam-break effect on a Solid Wall using \\
Smoothed Particle Hydrodynamic
\end{center}

\medskip

\begin{center}
{\sc Suprijadi$^{a,b}$, F. Faizal$^{b}$, C.F. Naa$^{a}$ and A.Trisnawan$^{b}$}

\medskip

$^{a}$Department of Physics, $^{b}$Department of Computational Sciences\\
Faculty of Mathematics and Natural Science, Institut Teknologi Bandung\\
Jl. Ganesha No.10, Bandung 40132, Indonesia\\
email : supri@fi.itb.ac.id
\end{center}

\bigskip

{\parindent 0pt
\textbf{Abstract.} {\em Dam is built for water supply, water flow or flooding control and electricity energy storage, but in other hand, dam is one of the most dangerous natural disaster in many countries including in Indonesia. The impact of dam break in neighbour area and is huge and many flooding in remote area, as happen in Dam Situ Gintung in Tangerang (close to Jakarta) in 2009. Smoothed Particle Hydrodynamics (SPH), is one of numerical method based on Lagrangian grid which is applied in astrophysical simulation may be used to solve the simulation on dam break effect. The development of SPH methods become alternative methods to solving Navier Stokes equation, which is main key in fluid dynamic simulation. In this paper, SPH is developed for supporting solid particles in use for 3D dam break effect (3D-DBE) simulation. Solid particle have been treated same as fluid particles with additional calculation for converting gained position became translation and rotation of solid object in a whole body. With this capability, the result of 3D-DBE simulation has been varies and interesting. The goals of this simulation is for analyse fluid and solid particle interaction by using two different scenario. The first scenario relation between height of fluid to a solid wall barrier and second scenario, is to study relation between solid wall and its collapse time by dam break. The results show sliding distance of a solid wall is depend on a fluid height, and relation between fluid heght and wall dimension will be discussed too in this paper.}
}


\section{Introduction}
Dam is built for water supply, water flow or flooding control and electricity energy storage, but in other hand, dam is one of the most dangerous natural disaster in many countries including in Indonesia. Because of the age, there are some evidence of natural disaster by a dam failure and break were reported for example in Situgintung  Tangerang, close to Jakarta was happen in 2009 \cite{bbc}. The impact of dam break in neighbour area and is huge and many flooding in remote area, the computer calculation to predict effect of the dam broken is important including its vast impact\cite{munger}.
 
Smoothed Particle Hydrodynamics (SPH), is one of numerical method based on Lagrangian grid which is applied in astrophysical simulation may be used to solve the simulation on dam break effect. The development of SPH methods become alternative methods to solving Navier-Stokes equation, which is main key in fluid dynamic simulation. As a standard test case in computational fluid dynamic the dam break simulation were often reported in many cases, including using SPH methods \cite{kelager}. The simulation using SPH became emerge whether using two or three dimension. In other hand, simulation of solid particle is many reported direct numerical simulation \cite{huang}. Interaction between fluid and solid particle is still open ended to studying. Including simulation base on SPH method.

In this paper, SPH is developed for supporting solid particles in use for 3D dam break effect (3D-DBE) simulation. Solid particle have been treated same as fluid particles with additional calculation for converting gained position became translation and rotation of solid object in a whole body. With this capability, the result of 3D-DBE simulation has been varies and interesting. The goals of this simulation is for analyse fluid and solid particle interaction by using two different scenario. The first scenario relation between height of fluid to a solid wall barrier and second scenario, is to study relation between solid wall and its collapse time by dam break. The results show sliding distance of a solid wall is depend on a fluid height, and relation between fluid height and wall dimension will be discussed too.

\section{Smoothed Particle Hydrodynamics}
The governing equation for fluid dynamics is Navier-Stokes equations consist of conservation of momentum, conservation of mass and conservation of energy. Assume the system of dam is in isothermal state, then we can neglect the calculation of fluid's thermal energy. The momentum equation has been written as follows:

\begin{equation}\label{eq1}
\rho\,\frac{\partial \mathbf{u}}{\partial t}+\rho\,\left(\mathbf{u}\cdot \nabla\,\mathbf{u}\right)=-\nabla p+\mu\nabla^2 \mathbf{v}+\rho\mathbf{g}
\end{equation} with $\rho$ is fluid density, t is time, \textbf {v} is fluid velocity, \textbf {p} is pressure, $\mu$ is viscosity, and \textbf {f} is external force. The fluid motion is also governed by the conservation of mass, which stated by
\begin{equation}\label{eq2}
\frac{\partial \rho}{\partial t}+\nabla \cdot \left(\rho\mathbf{u}\right) = 0
\end{equation}

Equations \eqref{eq1} and \eqref{eq2} will be solved by numerical approximation of  SPH method. The method simply described by discreet collection of particles or balls for approximating fluid body. Every particle mathematically influence the neighbouring particles via kernel function.
\begin{equation}\label{eq3}
W\left(r,l\right)=\frac{1}{l \sqrt{\pi}}\exp{\left(\frac{-r^2}{l^2}\right)}
\end{equation}
Here $W$ is kernel function, $x$ is particle distance and $l$ is smoothing factor. The kernel function stated in equation~\eqref{eq3} now can be use as weighting function to calculate all physical quantity, gradient and Laplacian attached to the particle, here we calculate the solution of equation~\eqref{eq1} and \eqref{eq2} using weighted interpolation concept and solved them numerically. The detail of this method described clearly in \cite{liu}.

\begin{figure}[!h]
\center \includegraphics[width=1.8in]{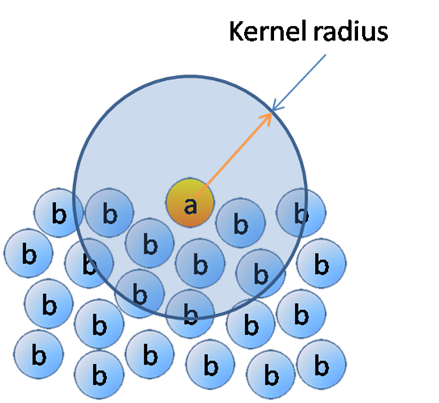}
\caption{Kernel radius in SPH simulation}
\label{fig1}
\end{figure}

Figure~\ref{fig1} illustrated the approximation of fluid body by collection of moveable particle interacted along kernel radius $r$. The numerical SPH for momentum and mass conservation int total derivative form respectively are \cite{monaghan}:
\begin{equation}\label{sphmoment}
		\frac{D\mathbf{u}_a}{Dt} =\sum_b m_b \left(\frac{p_b}{{\rho_b}^2}+\frac{p_a}{{\rho_a}^2} + \Pi_{ab}\right) \nabla_a W_{ab} + \mathbf{g}
\end{equation}
and
\begin{equation}\label{sphconti}
		\frac{D{\rho}_a}{Dt}=\sum_b m_b\,\left(\mathbf{u}_b-\mathbf{u}_a\right)\cdot \nabla_a W_{ab}.
\end{equation}
With index $a$ and $b$ denoted particle index, $\Pi_{ab}$ is numerical viscosity described in\cite{monaghan}. Here the kernel function $W_{ab}$ and it's derivative solved analytically. Equation~\eqref{sphmoment} and \eqref{sphconti} can be solved using direct numerical integration such as Euler method. Another well-known methods is Runge-Kuta method, Leap frog method and Verlet method.

\section{Solid - Fluid Interaction}
In DBE simulation, one of the important aspect is solid motion exerted by fluid energy. Here we use solid body dynamic to simulate the motion of fluid. Once the total force and total torsion exerted by fluid to surface of solid already known, so do translational and rotational motion of solid can be calculated. 
\begin{figure}[!h]
\center \includegraphics[width=5in]{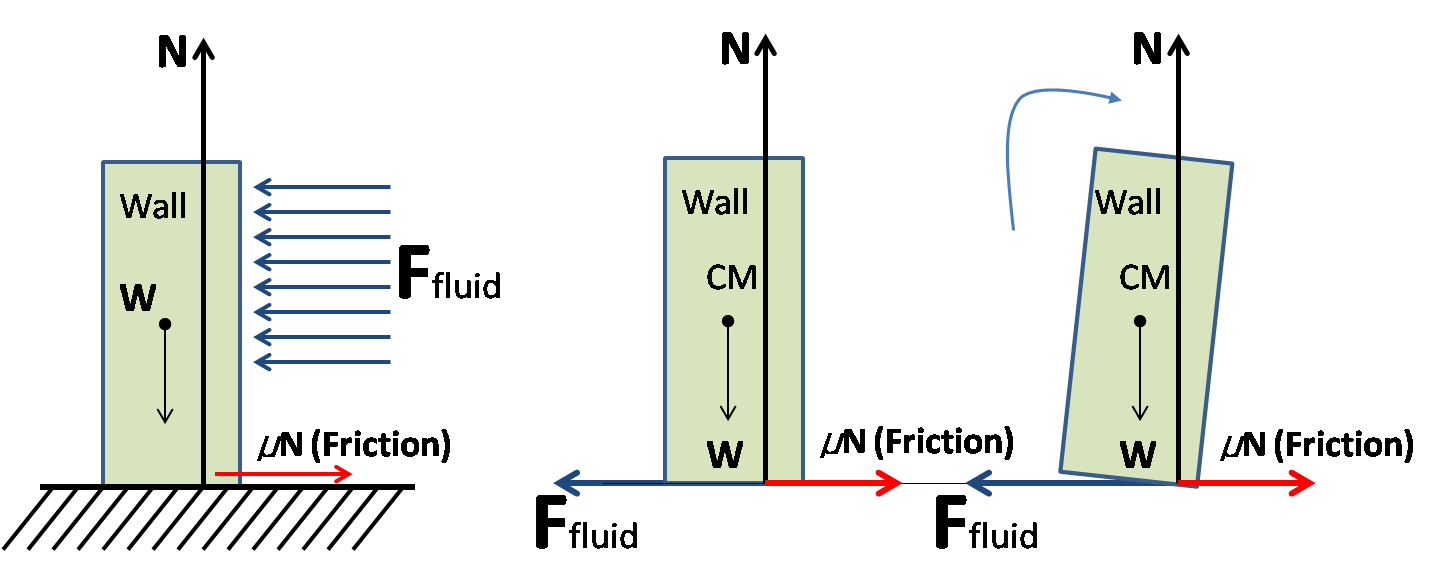}
\caption{Force diagram and solid wall motion}
\label{solidfluid}
\end{figure}
Figure~\ref{solidfluid} show the basic physical scheme of the solid fluid interaction with strong assumption that no flow across the surface of fluid particle. Some important aspect such as floor friction and center mass of solid body is also taking into account.

\begin{figure}[!h!]
\center \includegraphics[height=1.1in]{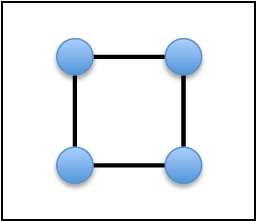}
\includegraphics[height=1.1in]{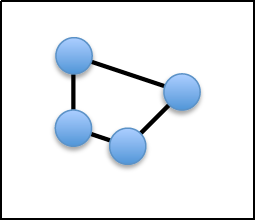}
\includegraphics[height=1.1in]{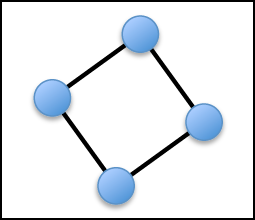}
\caption{Solid particle motion schematic}
\label{solidmot}
\end{figure}
To simplified the continuum aspect of solid body, solid object will be approximated also by collection rigid particles or balls with certain physical quantity governed by newton equation of translation and rotation. The center of mass of such discreet particle are also easy to calculate. Figure~\ref{solidmot} showed the step process of solid particle position update in rotational motion.
\section{Results and Discussion}
The scenario of DBE simulation in three dimension described as follow. Initially the fluid position in in cubic shape dam for simplification and without initial velocity of fluid assuming the fluid is at rest. At a certain distance lied solid wall to block flow of the fluid. Another assumption were made that there no crack or fracture in solid wall.
\begin{figure}[!h]
\center{\includegraphics[width=4in]{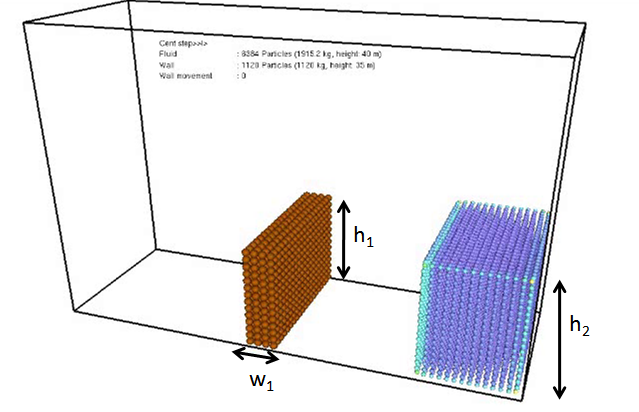}}
\center \caption{Simulation scenario diagram, $h_2$ is initial fluid height, $h_1$ and $w_1$ is height and width of solid}
\label{initial}
\end{figure}

Figure~\ref{initial} shown the description of initial arrangement of solid and fluid particles. As described earlier the arrangement of fluid particle in box shape dam and initially particle at rest. As the time step moves the gravity force work on the fluid particles and they move with the same for as if happened in the dam failure suddenly. 

Some parameter involved in 3D simulation in Figure~\ref{initial} is initial fluid height, solid wall dimension. Later in this section will be shown some result in 3D DBE simulation with some variation in simulation parameter.

\begin{figure}[htbp!]
\center \includegraphics[width=5.8in]{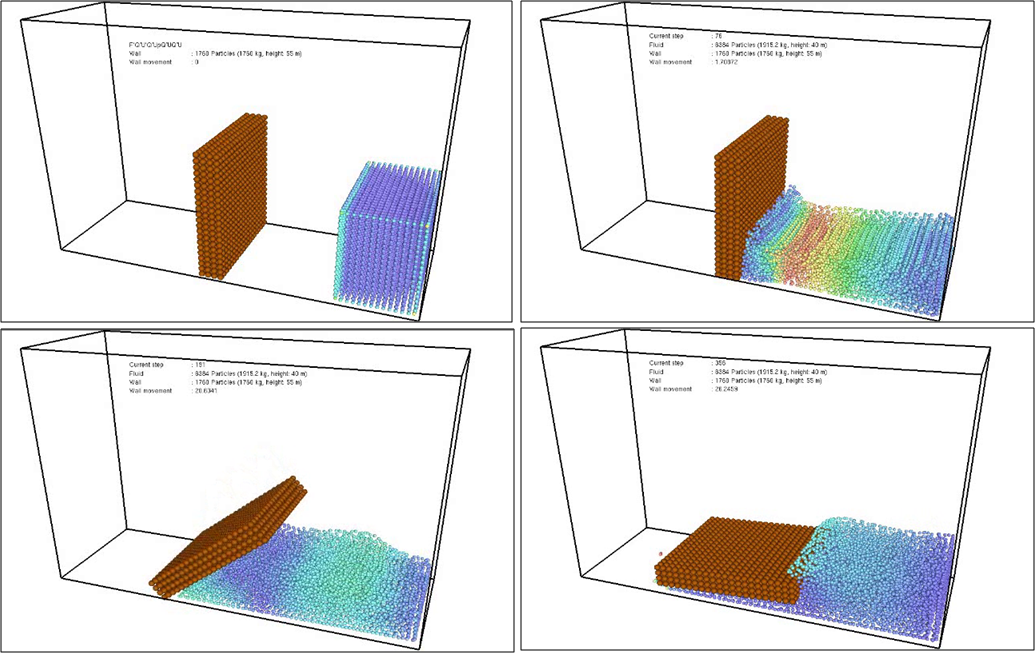}
\caption{3D simulation of DBE a $t_1=0 s$, $t_2=0.228 s$, $t_3=0.543 s$ and $t_4=1.068 s$, $h_1=55 m$ and $h_2=40 m$ and $w_1 =40 m$}
\label{3DDBE}
\end{figure}

Figure~\ref{3DDBE} show general result of DBE simulation wall with height $55 m$ and fluid dam with $40 m$ in height at time $t_1=0 s$, $t_2=0.228 s$, $t_3=0.543 s$ and $t_4=1.068 s$. The frame shows the fall process of the wall as the effect of fluid flow.

\begin{figure}[htbp!]
\center \includegraphics[width=5.8in]{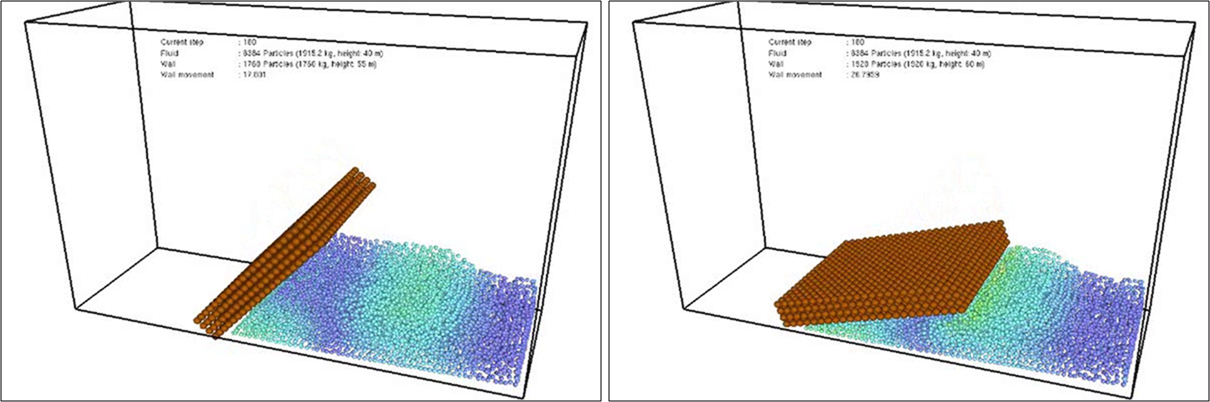}
\caption{Wall height effect on DBE simulation at $t= 0.54 s$. Left $h_1=55m$, right $h_1=60m$}
\label{wallheight}
\end{figure}

The variation in solid wall height is made to test the stability of moveable wall under the same energy of fluid flow exerted on the wall surface. It is done by setting the fluid height at $40 m$. Figure~\ref{wallheight} shows the effect of wall height to the fall process under the influence of flow caused by dam break event. Here the frame plot the position of particles at the same time step $0.54s$ and the same initial height of fluid $h_2=40m$ but in different wall height. The frames show different response of falling time. Higher wall fall faster than the lower one at other parameter were keep constant. Here we can say that the dimension of the wall influence the stability of the wall object under the same outer forces. If the width of the wall has been increasing the high wall also become relative stable.

In another scenario the variation of fluid height were tested to the relatively stable wall object and observe the wall sliding distance.
\begin{figure}[htbp!]
\center \includegraphics[width=4in]{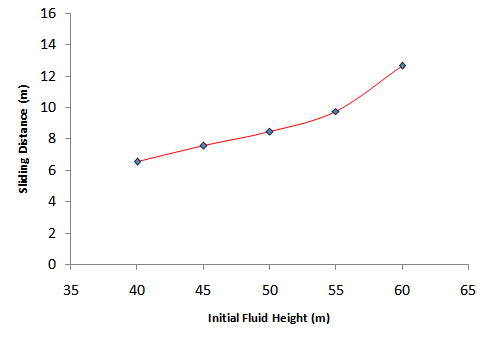}
\caption{Fluid height effect to the sliding distance on DBE simulation}
\label{fluidheight}
\end{figure}
Figure~\ref{fluidheight} shows sliding distance plot versus initial fluid height. The trend of the data seems to be non linear as the height of fluid increase over 55 m, this is make sense because of volume of the fluid increase by the power of three to the height measure and linearly increase the potential energy of fluid.

\section{Conclusion}
We successfully simulate 3D-DBE simulation using SPH methods, and fluid-solid interaction capability was applied to study the dam break effect on a solid wall object, the SPH can be used to simulate a solid particle simulation. Some simplified assumption that no crack or fracture happen in the solid wall object thus no separation of solid body as the impact of the fluid momentum. The simulation results showed that the fluid height and solid wall thickness and height play a significant effect on solid wall translational and rotational motion.

For the future work it is also challenging to study the fracture event, crumbling process or separation of solid body by the effect of momentum from the fluid flow and made realistic devastation process from dam failure. This features will increase the computational cost and difficulties in algorithm.

\section*{Acknowledgement}
A part of this research was funding by Research and Innovation Program of Research Group ITB, grant No:243/I.1.C01/PL/2013.

\end{document}